\documentclass[aps,prl,twocolumn,amsmath,amssymb]{revtex4}
\usepackage{dcolumn}
\usepackage{bm}
\usepackage{graphicx}
\begin{document}
\title{Overtaking while approaching equilibrium.}
\author{P. Chaddah, S. Dash, Kranti Kumar and A. Banerjee}
\affiliation{UGC-DAE Consortium for Scientific Research, University Campus, Khandwa Road, Indore-452001, Madhya Pradesh, India.}

\begin{abstract}
A system initially far from equilibrium is expected to take more time to reach equilibrium than a system that was initially closer to equilibrium. The old puzzling observation (also called Mpemba effect) that when a sample of hot water and another sample of cold water are put in a freezer to equilibrate, the hot water sometimes overtakes as they cool, has been highlighted recently. In the extensively studied colossal magnetoresistance manganites, cooling in a magnetic field (H) often results in an inhomogeneous mixture of transformed equilibrium phase and a kinetically arrested non-equilibrium phase which relaxes slowly towards equilibrium at fixed H and temperature (T). Here we show that the magnetization decay rate at the same H and T is larger for the state that was initially farther from equilibrium, and it continues to relax faster even after these have become equal. Our result should help propose an explanation, for Mpemba effect, that does not attribute it to any artifact. 
\end{abstract}

\maketitle
A system initially far from equilibrium is expected to take more time to reach equilibrium than a system that was initially closer to equilibrium. The counter-intuitive observation known `since the time of Aristotle' that hot water freezes faster (also called Mpemba Effect) has been highlighted recently\cite{ball} with the assertion ``it does seem as though hot water sometimes `overtakes' cold as they cool". In the extensively studied colossal magnetoresistance manganites, cooling in a magnetic field (H) often results in phase coexistence of a transformed equilibrium phase and a kinetically arrested non-equilibrium phase whose fraction R$_{NE}$ (directly related to the magnetization M) measures the separation from equilibrium\cite{ban1, cha, wu, ban2}. The approach to equilibrium (R$_{NE}$=0, or M=0) occurs on warming; the non-equilibrium state also relaxes slowly towards equilibrium at fixed H and temperature (T). Here we show the surprising result that a state with initially higher M  (and thus farther from equilibrium) overtakes a state with initially lower M (and thus closer to equilibrium) as they approach equilibrium. M values that are comparable relax with time t at drastically different rates (itself a surprising result), to the extent that their trajectories meet-and-cross rather than the intuitive meet-and-merge. This unexpected `overtaking' is reminiscent of the Mpemba Effect\cite{ball}. These results also have implications for the history-dependent coexistence of competing phases, which determines the functionality of various magnetic materials of current interest\cite{choi}. Our results on measurements that are free from artifacts and susceptible to rigorous experimental protocols, should help understand why the growth of the equilibrium phase is being influenced by how far the initial state was from equilibrium. 

The extensively studied half-doped manganite La${_{0.5}}$Ca${_{0.5}}$MnO${_{3}}$ (LCMO) shows, on cooling in magnetic fields, a 2nd order paramagnet-to-ferromagnet (FM) transition around 230K and a 1st order FM-to-antiferromagnet (AFM) transition around 150K\cite{ban1, cha, lou}. The regions of equilibrium (AFM) phase form down to about 100K on cooling in 1 Tesla. The sample used here is the same as used in our earlier studies\cite{ban1, cha}. Our earlier results for the zero-field cooled (ZFC), field-cooled cooling (FCC) and field-cooled warming (FCW) M-T curves in 1T field\cite{cha} indicated that the M-T behaviour is similar to that seen by Loudon et al.\cite{lou} At low-temperatures, many manganites exist as an inhomogeneous mixture of non-equilibrium (arrested) and equilibrium regions, with the transformation to equilibrium phase having been arrested below a T$_g$ that depends on the cooling field H$_{Cool}$\cite{ban1, cha, wu, ban2}. In half-doped manganites such as LCMO, the equilibrium low-T phase is AFM where the electronic charge is ordered whereas in the high-T FM phase electronic charge is a `liquid'. The disorder in the arrested FM state can be understood as that retaining the disorder of the charge-liquid\cite{ban3}. The kinetic arrest of a first order magnetic transition has been reported recently in many manganites\cite{ban1, cha, wu, ban2, choi, ban3, kranti, lak} and many other functional materials\cite{mkc, kus, sbr, yu, ito, san, ang}, and is referred to as a `magnetic glass'\cite{ban1, cha, wu, ban2, choi, ban3, kranti, lak, mkc, kus, sbr, yu}. This kinetically arrested `magnetic glass' is a metastable away-from-equilibrium state. How far from equilibrium is this state is measured by the fraction of the arrested FM phase, and is controlled by varying H$_{Cool}$. The fraction R$_{NE}$ of the FM phase, in the inhomogeneous mixture of FM and AFM regions, is related linearly to magnetization, and is a measure of how far the system is from equilibrium. (R$_{NE}$ at 5K is obtained by dividing M by 3.5, estimated assuming saturation magnetization of 3.5 $\mu_B$/formula unit).

Polycrystalline La${_{0.5}}$Ca${_{0.5}}$MnO${_{3}}$ sample has been prepared through a well-established chemical route known as `pyrophoric method'. Details of the sample preparation and characterization are given in reference\cite{cha}. The magnetic measurements are performed using commercial set-ups, SQUID magnetometer (MPMS XL, M/s. Quantum Design, USA). All the measurements are performed on the same piece of sample. To prepare states with different fractions of equilibrium phase, the sample is cooled from 320K to 5K in the different cooling field (H$_{Cool}$) at the same rate. At 5K, the field is changed to the respective measurement field and the data is taken while warming. For time dependent magnetization, each time, after producing states with different fractions of equilibrium phases, the sample is heated at the same rate from 5K to the measurement temperatures in the respective measurement fields. All the measurements are reproducible within the uncertainty, which is much less than the dimensions of data points depicted in the figures.        

On cooling to 5K in a magnetic field H$_{Cool}$, R$_{NE}$ varies from about 0.17 for H$_{Cool}$ = 0 to about 0.90 for H$_{Cool}$ = 6Tesla, and this fraction is retained when H is decreased isothermally at 5K since 5K is well below T$_g$ for all H in our range. We show in fig 1(a) magnetization measurements, with data taken on heating in 1Tesla, for H$_{Cool}$= 6 Tesla, H$_{Cool}$= 3 Tesla and H$_{Cool}$= 1 Tesla (to be denoted by C6T, C3T and C1T respectively).  M value is seen to reduce for C6T and C3T above 20K where nucleation of the equilibrium phase starts, reaching a shallow minimum value of M$_{min}$ around 100K. We recently reported \cite{ban1} the unexpected observation that if M was higher at 5K, then the corresponding M$_{min}$ is lower; fig 1(a) confirms this as M$_{min}$(C6T)$<$M$_{min}$(C3T)$<$M$_{min}$(C1T).  Thus M$_{min}$ is lower when R$_{NE}$ is higher, or the state at 100K is closer to equilibrium when the initial state at 5K is farther from equilibrium. We have confirmed the same behaviour by measurements of resistivity (which is related inversely but non-linearly to R$_{NE}$) under the same cooling and heating protocol [not shown here]. We have thus observed that a state initially farther from equilibrium overtakes a state initially closer to equilibrium as equilibrium is approached by raising temperature. We now come to the more interesting and puzzling results obtained by the basic procedure of approaching equilibrium by allowing the metastable state to relax isothermally, albeit at various temperatures. 

We note from fig 1(b) that the M for C6T and C3T become equal at 40.5K, with C6T overtaking C3T at higher temperatures in the approach to equilibrium. Similarly, M for C6T and C1T becomes equal at 56K, with C6T overtaking at higher temperatures in the approach to equilibrium. These crossover temperatures (T$_X$) are consistent with those obtained by resistivity measurements. In this range of temperatures, the relaxation to R$_{NE}$=0 equilibrium state is slow, and the time required for full conversion would exceed an experimenter's patience! There is nothing special about warming in a field of 1 Tesla, and we show similar crossovers in fig 1(c and d) for a warming field of 3 Tesla. Here C6T and C4T have a T$_X$ =38K, and C6T and C3T have a T$_X$ =54K. 

We now show the relaxation of magnetization at T$_X$=40.5K for C6T and C3T, and at T$_X$=56K for C6T and C1T cooling-histories for the 1Tesla measuring field. The results in fig 2 show a drastically higher isothermal decay of the FM regions at the same measuring field (1 Tesla) for C6Tin both cases even though the starting magnetization values are same in each cases. The decay for the C6T case can be fit to a power-law at long times, reminiscent of the report in electron glass\cite{amir}. (A similar behavior was observed in our resistivity measurements as well).

We now show time relaxation measurements at temperatures slightly below T$_X$. The results in fig 3(a) show the initially higher M state (C6T) `overtaking' (after about 35 minutes) the initially lower M state (C1T) as they relax towards equilibrium at 52.5K when measured in 1Tesla field. Similarly, for 3Tesla measuring field, we show in fig 3(b) the isothermal relaxation of M at 52K (T$<$T$_X$) for C6T and C3T. The initially higher M (and thus R$_{NE}$) in C6T relaxes faster (as R$_{NE}$ falls) than the M in C3T, and `overtakes' it after a relaxation time of about 45 minutes. The initially higher M in C6T relaxes faster than the M in the case of C2.9T state, and `overtakes' it after a relaxation time of 155 minutes as shown in fig 3(c). In fig 3(d) we show relaxations at a lower temperature of 37.5K where C6T overtakes both C4T (after 25 minutes) and C3.85T (after 75 minutes) as M falls in the approach to equilibrium.

Summarizing the data, we have two states that are inhomogeneous mixtures of the equilibrium AFM phase with a fraction R$_{NE}$ of the non-equilibrium FM phase, and the initially farther from equilibrium state approaches equilibrium faster. We now try to understand why the state with higher R$_{NE}$(5K) overtakes the state with initially lower R$_{NE}$(5K) state in the approach to equilibrium. We note that on heating the arrested state, dearest occurs, with a drop in M (or R$_{NE}$) and conversion of FM to equilibrium phase. This conversion (or nucleation of equilibrium regions) starts at a lower temperature when R$_{NE}$ (5K) is higher\cite{lak}. It is known that the nucleation temperature dictates the critical radius R$_C$ for nuclei formation\cite{lif, lif2, chaik}, and this is given by

\begin{eqnarray}
 R_C = \frac{2\sigma}{\Delta f}	
\end{eqnarray}

where  $\sigma$ is the `surface tension' associated with the interface,  and $\Delta$f is the difference in the bulk free energies of the two phases. Now $\Delta$f = 0 at the transition temperature T$_C$, and rises rapidly as T shifts away from T$_C$. Thus if nucleation of the low temperature phase takes place well below T$_C$, then regions of the AFM phase will have small radii, whereas if it takes place close to T$_C$ then the regions will have large radii.
      
We note, from figure 1(a) that for C6T warming in 1 Tesla, nucleation starts at 12 K, whereas for C3T warming in 1 Tesla it starts at 20 K. Similar behaviour is noted form figure 1(c) for warming in 3 Tesla. This clearly shows that R$_C$ would be lower when the initial R$_{NE}$ is higher. The fraction of equilibrium phase for two different R$_{NE}$(5K) is equal at T$_X$, but as argued above, the higher R$_{NE}$(5K) has a larger number of nuclei of smaller size. Thus the volumes of the equilibrium phase regions are equal at T$_X$, but the surface area is higher in the higher R$_{NE}$(5K) case. 

Relaxation of the coexisting-phase sample in manganites has been attributed to relaxation of the interfaces\cite{sha}, and we expect growth of existing nuclei, rather than further nucleation, to dominate isothermal relaxation at long times. We thus attribute the (much) higher isothermal relaxation to a much larger area of the interface of the initial state. This is consistent with the initially higher R$_{NE}$(5K) sample having larger number of nuclei, but of smaller size, contributing to much higher interface area.

The explanation offered above is based on our clear observation of nucleation starting farther from the 1st order transition temperature T$_C$ when the system is initially farther from equilibrium\cite{lak}, and consequently having a smaller R$_C$. The drastically higher relaxation rate observed by us for an initially farther-from-equilibrium state is novel and intriguing, and is likely to also be observed in other functional materials showing kinetic arrest of a 1st order magnetic transition \cite{choi}. Moreover, different values of R$_C$ would create nanostructures at different length-scales, having implications for their functionality. Does freezing hot water also result in ice nuclei of smaller size and large numbers than with cold water?  Specifically, ``hot water sometimes `overtakes' cold as they cool" questions Newton's law of cooling, whereas our results indicates that ``hot water sometimes `overtakes' cold as they freeze" could be understood and verified by studies on the nucleation and growth of ice.

\begin{figure*}
	\centering
		\includegraphics{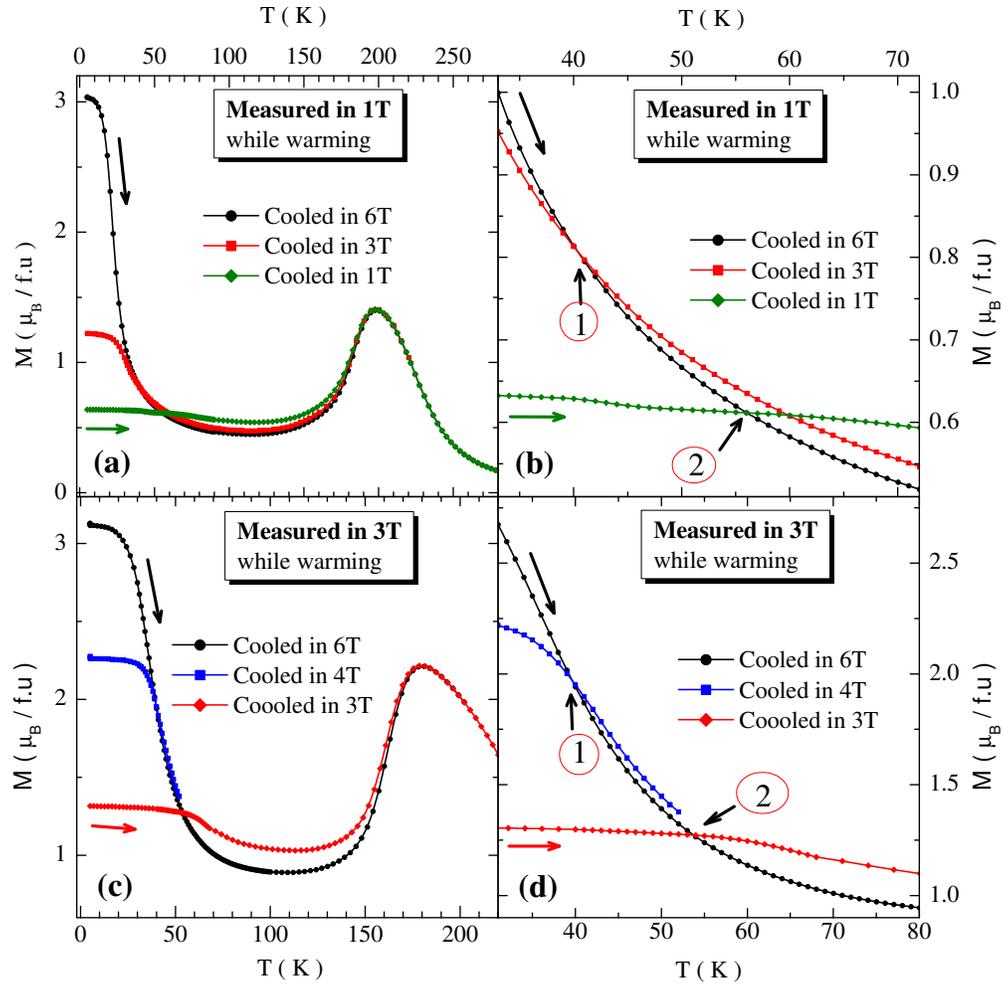}
	\caption{Magnetization as function of temperature while warming after cooling in different magnetic fields. (a) M vs. T while warming after cooling in 1, 3 and 6Tesla and isothermally changing the field at 5K to measurement field of 1Tesla. (b) Expanded view of a around the crossover. The crossover point between C6T and C3T is shown as (1) and that of C6T and C1T is shown as (2). (c) Shows M vs T following the similar protocol of a for 3Tesla measurement field. (d) Expanded view to show the crossover between C6T and C4T at  (1) and that of C6T and C3T at (2).}
	\label{fig:Fig1}
\end{figure*}

\begin{figure*}
	\centering
		\includegraphics{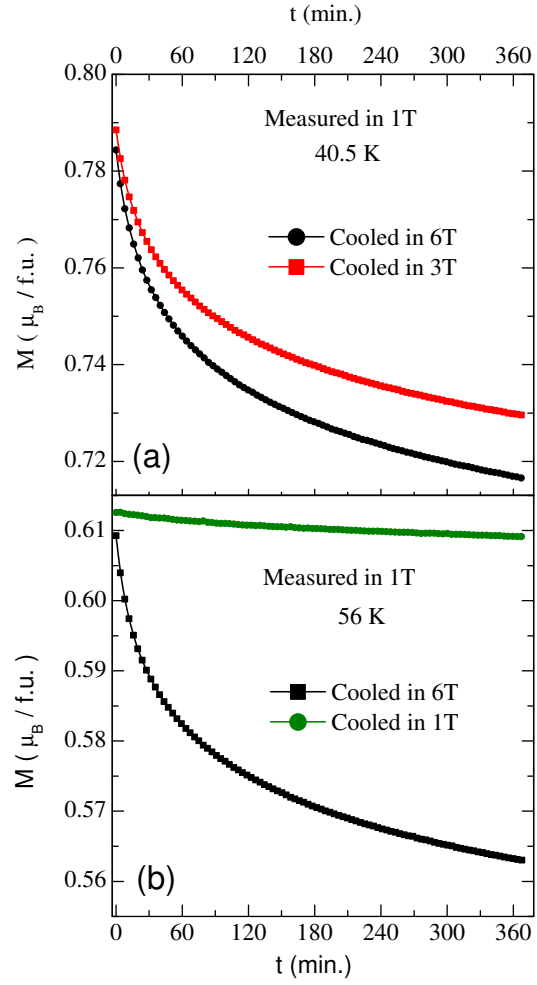}
	\caption{Approach to equilibrium by temporal relaxation: Magnetization relaxation in 1Tesla, measured after cooling in different field to 5K and isothermally changing the field to measurement field and warming the sample at a constant rate to the measurement temperature. (a) shows the relaxation of magnetization at 40.5K for C6T and C3T states. Note that though they start with the same value of M, they relax differently in the same field and temperature. (b) Shows drastically different rate of relaxation at 56K for C6T and C1T states when the starting M values are almost same.}
	\label{fig:Fig2}
\end{figure*}

\begin{figure*}
	\centering
		\includegraphics{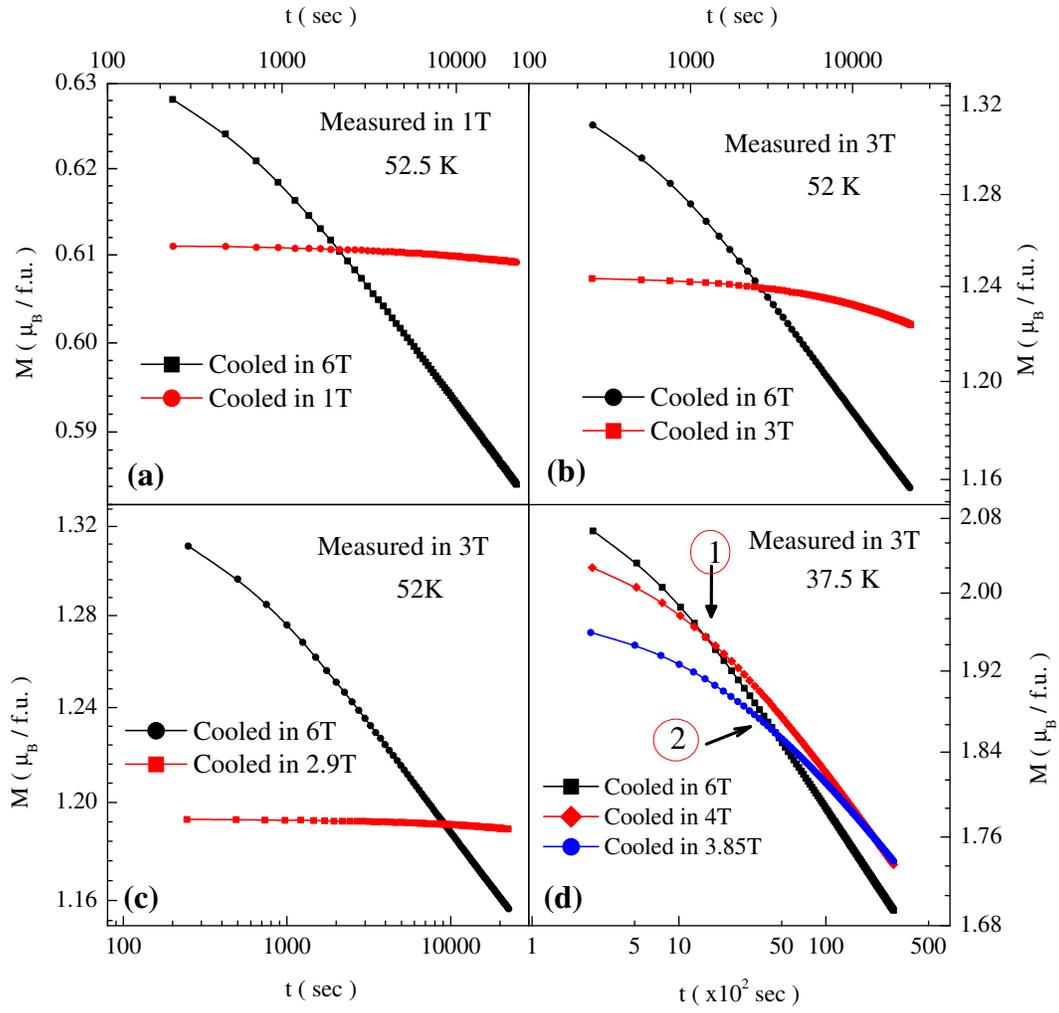}
	\caption{Overtaking while approaching equilibrium: Temporal relaxation of magnetization at the same fields and temperatures for different states prepared by cooling in different fields and isothermally changing to the measurement field at 5K. (a) Relaxation of M at 52.5K measured in 1Tesla field for C6T and C1T states. (b) Relaxation of M at 52K measured in 3Tesla field for C6T and C3T states. (c) Relaxation of M at 52K measured in 3Tesla field for C6T and C2.9T states. (d) Relaxation of M at 37.5K measured in 3Tesla field for C6T, C4T and C3.85T states showing how the states farther from equilibrium overtake the states closer to equilibrium while approaching equilibrium.}
	\label{fig:Fig3}
\end{figure*}

\end{document}